# Disentangling thermal birefringence and strain in the all-optical switching of ferroelectric polarization

Maarten Kwaaitaal*, Daniel Lourens, Carl S. Davies, and Andrei Kirilyuk*

FELIX Laboratory, Radboud University, Toernooiveld 7, 6525 ED Nijmegen, The Netherlands

*Corresponding authors: maarten.kwaaitaal@ru.nl; andrei.kirilyuk@ru.nl

## Abstract

Recent works have demonstrated that the optical excitation of crystalline materials with intense narrow-band infrared pulses, tailored to match the frequencies at which the crystal's permittivity approaches close to zero, can drive a permanent reversal of magnetic and ferroelectric ordering. However, the physical mechanism that microscopically underpins this effect remains unclear, as well as the precise role of laser-induced heating and macroscopic strains. Here, we explore how infrared pulses can simultaneously give rise to strong temperature-dependent birefringence and strain in ferroelectric barium titanate. We develop a model of these two coexisting effects, allowing us to use polarization microscopy to disentangle them through their spatial distributions, temporal evolutions and spectral dependencies. We experimentally observe strain-induced patterns that are an order of magnitude larger than that which can be accounted for by laser-induced heating alone, suggesting that non-thermal effects must also play a role. Our results reveal the distinct fingerprints of heat- and strain-induced birefringence, shedding new light on the process of all-optical switching of order parameters in the epsilon-near-zero regime.

## Introduction

Over the last half-century, a large body of research has aimed to discover how to rapidly and efficiently switch spontaneous polarization ordering from one non-volatile state to another. In the most standard approach, polarized domains are directly manipulated by the application of quasi-static or pulsed electric fields, but this does not conveniently facilitate faster (sub-nanosecond) switching speeds and spatial localization [1][1]. An alternative route comprises all-optical switching of ferroelectric polarization, in which ultrashort laser pulses are used to single-handedly drive the switching potentially across timescales on the order of picoseconds. Several all-optical schemes for achieving this have been proposed, based on impulsive Raman scattering **Error! Reference source not found.**, direct excitation of the soft ferroelectric mode [4], or non-linear phononics [7]. However, despite all these proposals, the experimental realization of laser-induced switching of polarization remains in its infancy. Recently, it was experimentally demonstrated that narrow-band optical pulses tailored to specific spectral lines in the infrared range can permanently switch ferroelectric domains in barium titanate ($BaTiO_3$) [10]. In particular, the switching becomes most pronounced when the pulse's frequency matches the frequency at which the material's complex permittivity is minimized i.e. in the epsilon-near-zero regime. Other works have demonstrated that an excitation within this regime also has the strongest effect when used to switch magnetization in ferrimagnetic iron-garnets and displace domain walls in antiferromagnetic nickel oxide [11]. The exact microscopic mechanism of this process, however, remains to be fully explained.

While experiments have demonstrated that infrared laser pulses can single-handedly switch polarization and magnetization, a question that always lingers relates to the inescapable impact of thermal effects. Research showed that laser heating can even be the sole cause of optical switching [14, 15]. Heating thus works in parallel with those induced by the structural strains induced via non-thermal effects, such as nonlinear phononics or infrared-resonant Raman scattering [8, 16]. Disentangling thermal and non-thermal effects is therefore imperative to understand the real underpinning mechanisms of the investigated phenomena.

Here, we numerically and experimentally investigate the combination of heating and accompanied strain effects in barium titanate upon excitation with infrared laser pulses. Using polarized light microscopy, we distinguish between crystal strains, which are revealed by changes in the birefringence stemming from photoelastic effects, and heating which can be observed due to a strong temperature-dependent birefringence in these crystals. These effects produce distinct symmetries which are observed *in-situ* during the actual process of all-optical switching of ferroelectric domains induced by the same excitation. We also show a strong correlation between the spectral dependencies of the optically switched domains from Ref. [10] and the measured strains here. Our measurements allow us to clearly isolate thermal and non-thermal effects induced by laser pulses in ferroelectric crystals.

## Results

### Simulations of birefringence- and strain-induced effects

Strong birefringence of $BaTiO_3$ in the tetragonal phase causes significantly dissimilar retardation of two orthogonally-polarized electromagnetic waves propagating through the crystal [16]. This retardation is proportional to the difference in the corresponding refractive indices ($n_i$) along the orthogonal crystal axes, with the total retardation ($\Delta\phi$) in a sample (of thickness $d$) being given by

$$\Delta\phi(T) = \frac{2\pi d \Delta n_{ij}(T)}{\lambda}, \qquad (1)$$

where $\lambda$ is the light wavelength and $\Delta n_{ij} = n_i - n_j$ the birefringence. Since the birefringence is a function of temperature ($T$), so is the retardation. The thermal dependence of birefringence under equilibrium conditions in $BaTiO_3$ single crystals is well-known[17].

Let us now assume the $BaTiO_3$ crystal is oriented such that its optical axes define the *x*- and *y*-axes. The crystal is illuminated with a linearly-polarized probing laser pulse (of wavelength 520 nm), with the latter's electric field being oriented at 45° relative to the *x*-axis. Upon irradiation with a pumping spatially-Gaussian laser pulse, the crystal experiences heating that presumably follows the spatial profile of the focused pump. The temperature profile at the sample is thus given by

$$T(r) = T_0 + \Delta T \cdot \exp\left(-\frac{r^2}{a^2}\right), \qquad (2)$$

where $T_0$ is the crystal's starting temperature, $\Delta T$ is the change in temperature induced by the pulse, and $a$ is the pulse's radius. The inhomogeneous heating of the sample leads to a similarly inhomogeneous variation of the crystal's birefringence. Using Eqs. (1)-(2) in combination with the temperature-dependent birefringence measured for $BaTiO_3$ [17], we calculate the retardation introduced in a 0.5-mm-thick crystal by a Gaussian pulse of heat of amplitude $\Delta T \approx 25$ K (Fig. 1a). Despite this temperature increase being rather modest, the rather strong birefringence of $BaTiO_3$ leads to several full-cycle rotations of the incident light's linear polarization (Fig. 1b). The variation in retardation can then be detected by transmitting the linearly-polarized light through the crystal and subsequently a polarizer ("analyzer"). This type of filtration, with the analyzer's transmission axis being inclined at -45° relative to the *x*-axis, gives rise to an oscillating pattern of measured intensity as shown in Fig. 1c, with the contrast being proportional to $\sin^2(\Delta\phi/2)$. In two dimensions, the radial symmetry of the laser-induced heating gives rise to a set of concentric rings (Fig. 1d). By therefore counting the number of rings, one can quite accurately estimate the temperature rise induced by the laser pulse within the sample.

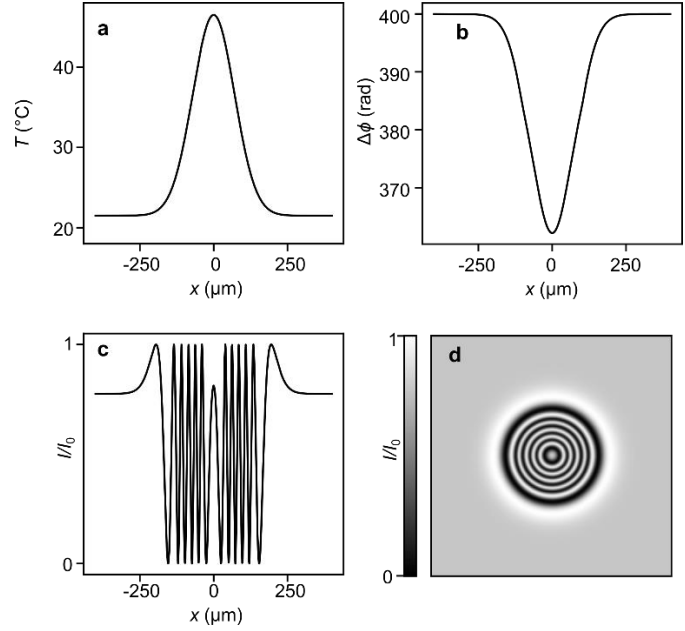

**Fig. 1** Calculated one-dimensional profile of **(a)** the thermal load delivered by a Gaussian laser pulse, **(b)** the resulting change in the birefringence, and **(c)** light intensity upon transmission through an analyzer. **(d)** The same profile is shown in panel (c), but in two dimensions.

In addition to the temperature-dependent effects on birefringence, the local heating delivered by a Gaussian-shaped pulse leads to an expansion of the affected area. In an isotropic material subject to thermal expansion as described by Eq. (2), the spatial distribution of the strain tensor *S* is given by [11,18]

$$S_{xx}(x, y) = \frac{\beta}{x^2 + y^2}\left[\left(1 - \exp\left(-\frac{x^2 + y^2}{2a^2}\right)\right)\left(1 - \frac{2x^2}{x^2 + y^2}\right) + \frac{x^2}{a^2}\exp\left(-\frac{x^2 + y^2}{2a^2}\right)\right], \quad (3)$$

$$S_{yy}(x, y) = \frac{\beta}{x^2 + y^2}\left[\left(1 - \exp\left(-\frac{x^2 + y^2}{2a^2}\right)\right)\left(1 - \frac{2y^2}{x^2 + y^2}\right) + \frac{y^2}{a^2}\exp\left(-\frac{x^2 + y^2}{2a^2}\right)\right], \quad (4)$$

$$S_{xy}(x, y) = -\frac{\beta xy}{x^2 + y^2}\left[\frac{2}{x^2 + y^2}\left(1 - \exp\left(-\frac{x^2 + y^2}{2a^2}\right)\right) - \frac{1}{a^2}\exp\left(-\frac{x^2 + y^2}{2a^2}\right)\right], \quad (5)$$

where $\beta = \frac{qa^2\Delta T}{3}\cdot\frac{1+\sigma}{1-\sigma}$, *q* is the thermal expansion coefficient and *σ* is the Poisson ratio of the material. The spatial distributions of the strain components $S_{xx}$ and $S_{xy}$ are visualized in Fig. 2(a)-(b) respectively. The spatial distribution of $S_{yy}(x,y)$ is equivalent to a rotation of $S_{xx}(x,y)$ by 90°. In general, this laser-induced strain influences the refractive indices of the irradiated material via the photoelastic effect. This effect is often described in terms of a change in the relative dielectric impermeability tensor ($B = 1/n^2$). By neglecting higher-order terms, we arrive at the commonly used expression for the effect of strain on the impermeability [19]:

$$\Delta B_{ij} = \rho_{ijkl} S_{kl}, \tag{6}$$

where $p_{ijkl}$ are the elasto-optic coefficients of BaTiO$_3$ [20]. As a consequence of BaTiO$_3$ being of point group 4mm and assuming a 2D strain profile, $B_{ij}$ will obtain the following form:

$$B_{ij} = B_{ij}^0(T) + \Delta B_{ij} = \begin{pmatrix} B_{xx} & B_{xy} & 0 \\ B_{xy} & B_{yy} & 0 \\ 0 & 0 & B_{zz} \end{pmatrix}, \tag{7}$$

$$B_{xx} = B_{xx}^0(T) + \rho_{xxxx} S_{xx} + \rho_{xxyy} S_{yy}, \tag{8}$$

$$B_{yy} = B_{yy}^0(T) + \rho_{yyxx} S_{xx} + \rho_{yyyy} S_{yy}, \tag{9}$$

$$B_{xy} = \rho_{xyxy} S_{xy}, \tag{10}$$

with $B_{ij}^0(T)$ the unstrained, temperature-dependent impermeability and $\Delta B_{ij}$ the strain-induced changes.

It is insightful to understand the effect of the different components of the strain tensor. The longitudinal components ($S_{xx}$ and $S_{yy}$) contribute to the diagonal terms of the refractive index and thus affect the magnitude of the birefringence along the strain profile. As a result, the pattern of concentric rings shown in Fig. 1d becomes deformed when including the strain within the simulations. The amplitude of this deformation is rather small when using the strain profile obtained from thermal expansion [Eqs. (3)-(5)], causing the final pattern to be broadly unchanged. To therefore highlight the effect of the strain, we present in Fig. 2c the pattern we obtain when artificially increasing the thermal expansion coefficient by a factor of 10.

In addition to the effect of the longitudinal components of the strain tensor, the transverse component $S_{xy}$ also changes the off-diagonal terms of the birefringence, and as a result induces a rotation of the optical axes in the material. The effect of the latter rotation is very small compared to the effect of thermally-induced birefringence. To therefore make this visible, one can orient the incident light polarization so it is collinear with one of the crystal's optical axes, which suppresses the contrast of the concentric rings (see Fig. 2d). For the simulations, the impermeability changes can be converted back again to a strain and temperature-dependent refractive index (see methods).

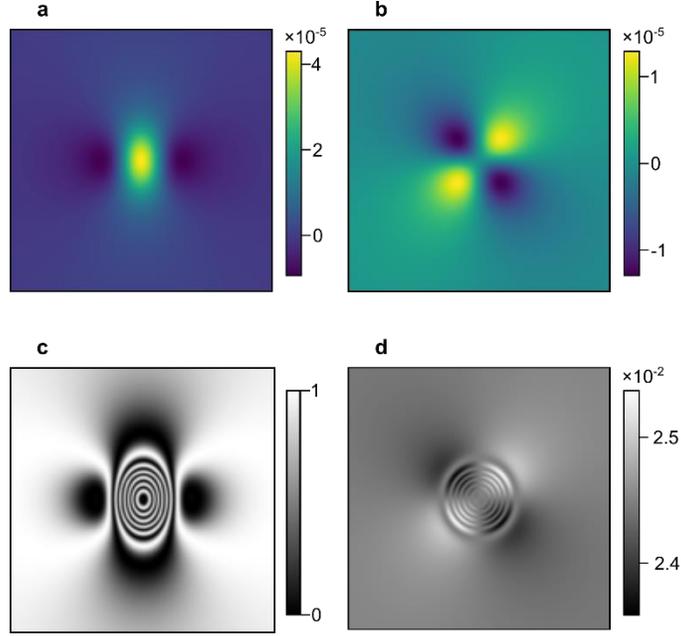

**Fig 2. (a)-(b)** Calculated spatial distribution of the strain components $S_{xx}$ and $S_{xy}$ respectively, resulting from laser-induced thermal expansion as described by Eqs. (3) and (5). **(c)** Calculated change in (normalized) light intensity induced in BaTiO$_3$ by both thermal birefringence and strain. Note the strength of the latter has been artificially amplified by increasing the thermal expansion coefficient by a factor of 10. **(d)** Calculated change in light intensity as detected with an analyzer, obtained with the input linear polarization aligned parallel to one of the crystallographic axes.

## Experimental measurements

To test how the birefringence and strain effects appear in practice, we perform multi-timescale pump-probe experiments using the intense infrared pulses delivered by the FELIX facility in the Netherlands (see Methods). In the first experiment, we pump BaTiO$_3$ with a single burst of ≈200 pulses, with a central wavelength of 8 μm, and probe with CW laser light with a temporal resolution of 27 μs (defined by the minimum exposure time of the camera). Upon acquiring an image immediately after the departure of the macropulse (at *t* = 8 μs [Fig. 3a]), we observe a rich combination of features including concentric rings and lobes along the diagonals. By electronically scanning the time at which the camera records an image, we obtain temporal resolution, with example images measured at further times shown in Fig. 3a-d.

Let us discuss the concentric rings, which are straightforwardly explained in terms of the thermally-induced birefringence (Fig. 1d). On a millisecond time scale, these rings gradually disappear one by one and the overall effect becomes more spatially spread out. Moreover, by analyzing the intensity profile of the rings, we can estimate the maximum retardation ($\Delta\phi$) at the center of the laser spot, which gives us the corresponding increase of the sample temperature. In Fig. 3e, we plot the measured reduction of the maximum retardation over time, along with the corresponding estimate of the crystal's temperature. By solving the equation for heat diffusion in the two-dimensional plane (Supplementary Information), we obtain a generalized temperature dependence $\Delta T \propto (1 + cT)^{-1}$ at the center of the irradiated spot of the form, where *c* is a proportionality constant. This equation fits well the experimentally-measured temporal dependence of the retardation (red dashed line in Fig. 3e). Thus, based on both the symmetry of the rings and their decay in time, we can conclude that the concentric rings are of purely thermal origin.

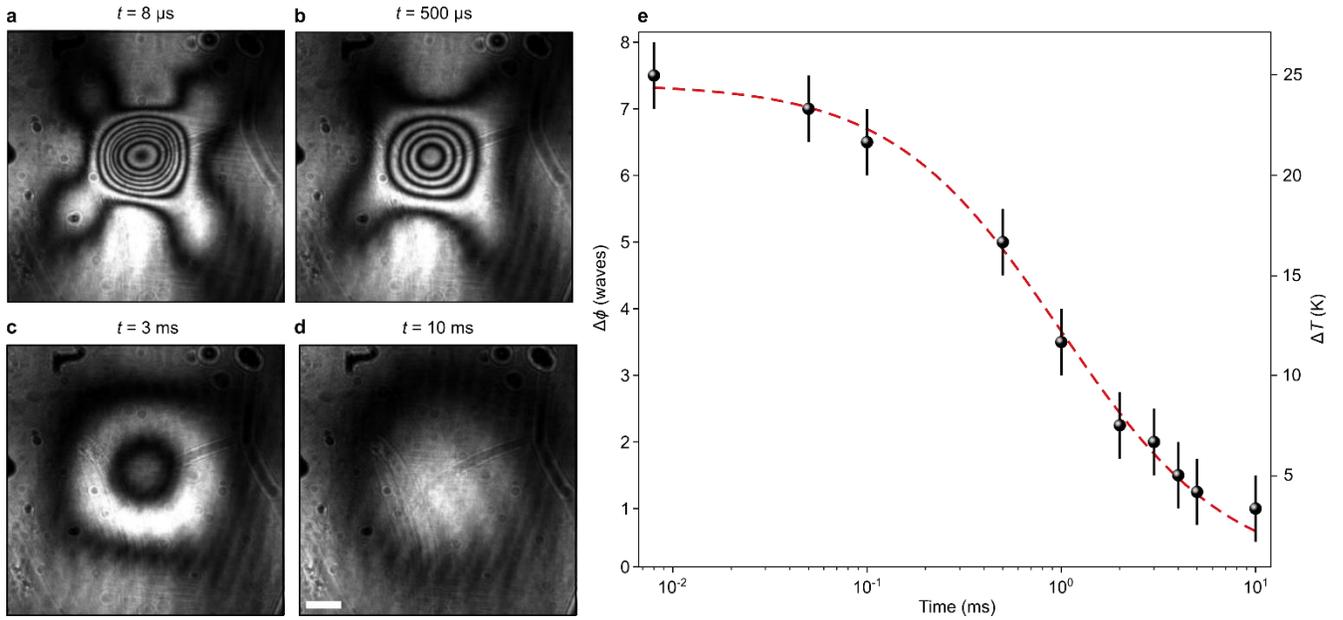

**Fig 3. (a)-(d)** Images taken using polarization microscopy after an excitation with an 8-μs-long burst of 200 pulses of wavelength $\lambda$ = 8 μm. The scale bar (common to all images) is 100 μm. **(e)** The time-varying birefringence observed in BaTiO$_3$ when pumped by a 10-ns-long burst of 10 micropulses. The red dashed line corresponds to a fit obtained by considering thermal diffusion in the sample plane.

At the longer timescales [Fig. 3(c)-(d)], the concentric rings are circular in shape. However, at the shorter timescales, they are rather rectangular. This does not stem from any ellipticity of the incident infrared pulses, as we verified that the latter are indeed Gaussian. Moreover, we rotated the sample while continuously irradiating, and observed that this pattern followed the rotation exactly. Thus, we conclude that the asymmetry originates from the crystal structure. In addition, we observe lobes along the diagonals, extending beyond the concentric rings. These lobes decay faster than the concentric rings. Since the spatial symmetry of the lobes is similar to that of the off-diagonal components $S_{xy}$ of the strain tensor (Fig. 2b), it is reasonable to infer that the lobes rather originate from strain.

To explore the importance of the pumping wavelength, we varied the central wavelength of the infrared pulses within the spectral range 8-24 μm (417-1250 cm$^{-1}$). In the top row of Fig. 4, we present typical images recorded with different pump wavelengths, obtained in the same way as those shown in Fig. 3. For $\lambda$ = 10 μm (Fig. 4a), we observe both the concentric rings and lobes along the diagonals and an additional pair of lobes along the horizontal axis. The lattermost feature was observed in the simulations (Fig. 2d) as well. Upon increasing the wavelength, we observe that the number of concentric rings decreases, the lobes along the diagonals become weaker, and the lobes along the horizontal axis vanish.

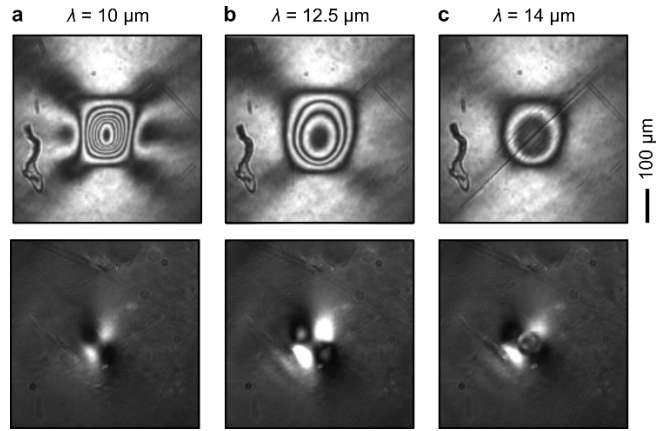

**Fig 4. a-c** Images taken upon pumping BaTiO$_3$ at the wavelength indicated. Top (bottom) panel: images taken with the electric field of the incident probing light polarized intermediate between (collinear with one of) the optical axes of BaTiO$_3$ The images in the top (bottom) panel were taken immediately after the departure of the pump excitation, which consists of a 8-μs-long burst of 200 pulses (10-ns-long burst of 10 pulses). The scale bar is common to all images.

To investigate the pattern further, we next oriented the transmission axis of the polarizer parallel to one of the optical axes of BaTiO$_3$. This configuration suppresses the birefringence, allowing us to see the much fainter optical effects from the shear strain $S_{xy}$. The visible four-lobe pattern is shown in the bottom panels of Fig. 4. We observe four lobes forming a quadrupolar pattern, with dark (bright) contrast on the [1,-1] (1,1) diagonals. These patterns can be compared directly to the simulation (Fig. 2c).

We carefully measured both temperature changes and strain amplitudes as a function of the excitation wavelength. In Fig. 5 we plot, as a function of the wavelength, the obtained strain amplitudes together with the maximum temperature increase, as given by the number of rings. While ΔT peaks at shorter wavelengths, strain is maximized in a different part of the excitation spectrum, thus demonstrating that we can indeed disentangle these effects.

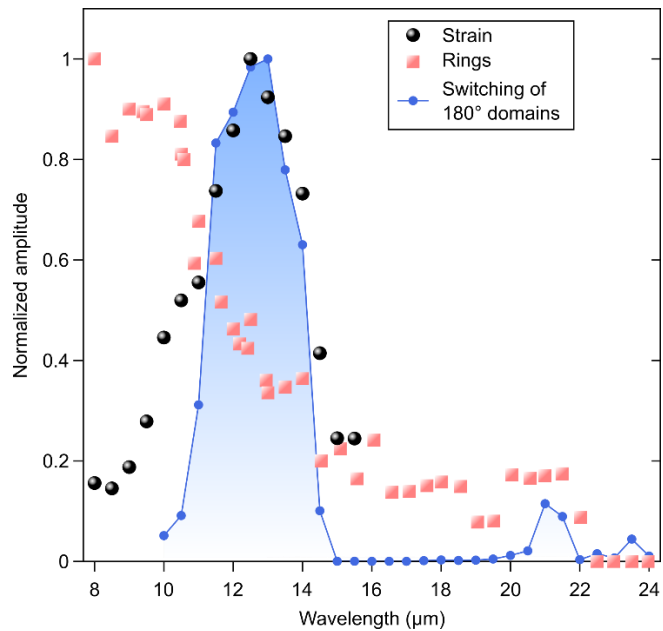

**Fig. 5** Spectral dependence of the temperature increase $\Delta T$ induced in BaTiO$_3$, as given by the number of concentric rings, and the magnitude of the strain, as given by the intensity change from the quadrupolar pattern. Also overlaid is the spectral dependence of the area of switched domains, taken from Ref. [10]. All values are normalized with respect to the incident pulse energy.

## Discussion

In Ref. [10], it was shown that ferroelectric polarization in barium titanate can be permanently switched upon pumping with narrow-band infrared pulses. To drive the switching, the pulses needed to be tuned in frequency to match the frequencies at which the dielectric constant goes close to zero. To therefore identify what precisely drives the switching, we overlay in Fig. 5 the normalized spectral dependence of the switching of 180° domains, measured in Ref. [10]. We clearly observe that the spectral dependence of the strain closely follows that of the switching of 180° domains. This result underscores the model used in Ref. [10] to explain the polarization switching in terms of laser-induced strain. Moreover, such radial strain pattern, as assumed by our model and confirmed by the experiments, is in excellent agreement with the model proposed in Ref. [20] to describe the magnetization pattern obtained in a canted antiferromagnet FeBO$_3$ by a similar phononic mechanism. This observation once again indicates the universality of the mechanism for switching of the order parameters in solids.

While the experimentally-measured patterns can be qualitatively explained by the simulations, we find it impossible to quantitatively reproduce the strain-induced pattern using realistic parameters of BaTiO$_3$. The image shown in Fig. 2d represents the best type of numerical reproduction we could obtain, explaining the appearance of the two side lobes. However, the symmetry of the pattern is not completely reproduced. More critically, we also needed to artificially increase the strain directly expected from thermal expansion by a factor of 10, in order to obtain a picture that most closely resembles the experimental observation (shown in e.g. Fig. 4a). We therefore conclude that thermally-induced strains are not capable of explaining the all-optical switching of ferroelectric polarization demonstrated in Ref. [10]. The amplification of the strain can reasonably be explained by non-thermal enhancement of the penetrating electric field via the epsilon-near-zero condition.

A temperature-dependent birefringence is not a special property of barium titanate, it is seen in many ferroelectric crystals. Therefore we suspect that the heat-rings can also be found in different ferroelectrics and could be used in other laser experiments involving ferroelectrics, acting as a thermometer.

## Methods

**Materials**

The BaTiO$_3$ crystals were commercially obtained from MSE Supplies. The crystals, of size 5 × 5 × 0.5 mm$^3$, were double-side polished and have a (100) crystal orientation. As a result, the ferroelectric polarization aligns along the surface (forming so-called a-domains).

**Numerical calculations**

*Jones calculus*

We use Jones matrices to evaluate how the intensity of probing light transmitting through the system is affected. The polarization of the incident light $\vec{E_i}$, after passing through the sample and analyzer, is measured to be in a final state $\vec{E_f}$ given by:

$$\vec{E}_f = \mathbf{J}_A \cdot \mathbf{J}_S \cdot \mathbf{J}_P \vec{E}_i, \tag{11}$$

where **J**$_A$, **J**$_P$ and **J**$_S$ are the Jones matrices of the analyzer, polarizer, and sample respectively. With both the transmission axis of the polarizer/analyzer and the optical axis of the crystal being inclined at $\vartheta$ relative to the x-axis, the Jones matrices **J**$_A$ and **J**$_P$ are given by

$$\mathbf{J}_{A,P} = \begin{pmatrix} \cos^2\theta & \cos\theta\sin\theta \\ \cos\theta\sin\theta & \sin^2\theta \end{pmatrix}. \tag{12}$$

The Jones matrix **J**$_S$ is defined by

$$\mathbf{J}_S = \begin{pmatrix} [\cos^2\theta + \sin^2\theta]\exp(i\Delta\phi) & [1-\exp(i\Delta\phi)]\cos\theta\sin\theta \\ [1-\exp(i\Delta\phi)]\cos\theta\sin\theta & [\cos^2\theta + \sin^2\theta]\exp(i\Delta\phi) \end{pmatrix} \exp\left(-\frac{i\Delta\phi}{2}\right), \tag{13}$$

Where $\Delta\varphi$ is the phase difference between the two orthogonal polarization components of the light, as given by Eq. (1). Using Eq. (11), we can obtain a spatially-resolved profile of light intensity $I$, since the latter is proportional to the square of the electric field $I \propto |E_f|^2$.

The magnitude of the birefringence and orientation of the optical axis may vary with position and temperature in the sample. To find these values for a given strain and temperature profile, we diagonalize the impermeability tensor defined in Eq. (6). Since the tensor is symmetric, the diagonalization can be achieved using the Mohr circle construction [19]. As a result, one obtains a diagonalized matrix $B'_{ij}$ by rotating the crystal's optical axis by an angle $\alpha$:

$$B'_{i,j} = \begin{pmatrix} \dfrac{B_{xx} + B_{yy} + \sqrt{(B_{xx}-B_{yy})^2 + 4B_{xy}^2}}{2} & 0 & 0 \\ 0 & \dfrac{B_{xx} + B_{yy} - \sqrt{(B_{xx}-B_{yy})^2 + 4B_{xy}^2}}{2} & 0 \\ 0 & 0 & B_{zz} \end{pmatrix}, \tag{14}$$

where $2B_{xy} = (B_{yy} - B_{xx})\tan 2\alpha$.

*Solution of two-dimensional time-dependent heat equation*

To describe the temporal evolution of the laser-induced retardation experimentally measured in Fig. 3e, we consider the heat diffusion equation. To make this problem analytically solvable, we neglect boundary conditions at the crystal surfaces and model the given problem as a two-dimensional heat transport problem. In particular, the equation is of the form

$$\begin{cases} \dfrac{\partial T}{\partial t} - D\nabla^2 T = 0 \\ T(x)|_{t=0} = g(x) \end{cases}, \tag{15}$$

where $g(\vec{x})$ is the initial Gaussian temperature profile at $t = 0$ given by Eq. (1). The solution to this problem is well-known (see e.g.[22]), being given by

$$\begin{cases} T = \int \Phi(\vec{x}-\vec{y})g(\vec{y})d\vec{y} \\ \Phi(\vec{y}) = \dfrac{1}{4\pi Dt}\exp\left(\dfrac{\vec{y}\cdot\vec{y}}{4\pi Dt}\right) \end{cases}, \tag{16}$$

The function $\Phi(\vec{y})$ is also known as the heat kernel. Combining these yields

$$T(\vec{x},t) = \frac{\Delta T}{4\pi Dt} \iint \left[ \exp\left( -\frac{(x_1-y_1)^2 + (x_2-y_2)^2}{4Dt} \right) \right] \cdot \left[ \exp\left( -\frac{y_1^2 + y_2^2}{2a^2} \right) \right] dy_1 dy_2, \tag{17}$$

Since we are primarily interested in evaluating the temperature at the center of the spot, we can simplify the integral by setting $x_1 = x_2 = 0$ and transforming from Cartesian to polar coordinates, yielding the solution:

$$T(t) = \frac{\Delta T}{1 + \frac{4Dt}{2a^2}}. \tag{18}$$

**Experimental measurements**

In the experiments, the BaTiO$_3$ crystals were exposed to infrared radiation provided by the free-electron lasers at the FELIX facility in Nijmegen, the Netherlands. The transform-limited infrared laser pulses ("micropulses") delivered by FELIX come at a rate of either 25 MHz or 1 GHz, and are conventionally enveloped within 10-μs-long bursts ("macropulses") arriving at a rate of 10 Hz. The micropulses have a duration on the order of 1-3 ps, a bandwidth of 0.3-1 %, and an energy on the order of 10 μJ. The laser beam was focused at normal incidence to a spot with a diameter of about 200 μm onto the surface of the BaTiO3 crystal.

The BaTiO$_3$ samples were probed using a polarization-sensitive microscope, with illumination provided by either a continuous-wave diode laser of wavelength 520 nm (FWHM spectral linewidth of 1 nm) or an amplified Ti:sapphire laser delivering 25-fs-long pulses of wavelength 780 nm (bandwidth 80 nm). These sources deliver light with linear polarization, with its orientation being adjustable by a half-waveplate. Upon transmitting through the BaTiO3 crystal at normal incidence, the light is collected by a ×10 objective lens and directed through a rotational polarizer ("analyzer"), before being captured by a CMOS camera (Thorlabs CS235MU). The sample is mounted on a rotational stage, allowing the sample's crystal axes to be rotated with respect to the incident electric field and the transmission axis of the analyzer.

The time resolution of our experiments was defined by the effect under investigation. In the first case, when studying the thermally-induced birefringence effects, we used the continuous wave laser to illuminate the sample and exposed the BaTiO$_3$ crystal to a single macropulse containing pulses coming at 25 MHz. A suitably-delayed electronic trigger was used to activate the exposure of the CMOS camera, with the exposure time set to 27 μs. This proved sufficient to study the thermally-induced birefringence effects, but insufficient to resolve the strain effects. In the second case, when studying the strain effects, we used the 25-fs-long pulses delivered by the amplified Ti:sapphire laser system for illumination. Moreover, we used the technique of "pulse-slicing" to obtain a pump excitation consisting of a burst of ≈10 micropulses separated in time by 1 ns. An electronic trigger was used to shift the temporal overlap between the pump and probe.

## Acknowledgements

We thank the technical staff at FELIX for providing technical support. D.G.L. acknowledges funding by the Max Planck–Radboud University Center for Infrared Free Electron Laser Spectroscopy. A.K., M.K. and C.S.D. acknowledge funding by the Nederlandse Organisatie voor Wetenschappelijk Onderzoek (Netherlands Organisation for Scientific Research). C.S.D. acknowledges support from the European Research Council ERC grant agreement no. 101115234 (HandShake).

## Author contributions

A.K. conceived the project. M.K. and D.G.L. performed the measurements. C.S.D. assisted in building the experimental set-up. A.K, M.K and C.S.D. wrote the paper. All authors were involved in discussing the results.

## Data availability statement

The data supporting the findings of this study are available within the article and its Supplementary Information. Additional data can be obtained from the authors upon a reasonable request.

## Additional information

**Competing interests**

The authors declare no competing interests.

**Corresponding authors**

M. Kwaaitaal (maarten.kwaaitaal@ru.nl) & A. Kirilyuk (andrei.kirilyuk@ru.nl)